\documentclass[useAMS,usenatbib,usegraphicx]{mn2e}

\title[The OLS--lens survey: The discovery of five new galaxy--galaxy
strong lenses from the SDSS]{The OLS--lens survey: The discovery of
five new galaxy--galaxy strong lenses from the SDSS\thanks{Based on
observations made with the William Herschel Telescope operated on the
island of La Palma by the Isaac Newton Group in the Spanish
Observatorio del Roque de los Muchachos of the Instituto de
Astrofisica de Canarias, and at the European Southern Observatory, Chile
(programme ID: 075.A-0463).}}  \author[J. P. Willis, et al.]
{J. P. Willis$^{1}$\thanks{E-mail: jwillis@uvic.ca}, P. C.
Hewett$^{2}$, S. J. Warren$^{3}$, S. Dye$^{4}$ and N. Maddox$^{2}$\\
$^{1}$Department of Physics and Astronomy, University of Victoria,
Elliot Building, 3800 Finnerty Road, Victoria, V8V 1A1, BC, Canada\\
$^{2}$Institute of Astronomy, Madingley Road, Cambridge, CB3 OHA, UK\\
$^{3}$Blackett Laboratory, Imperial College of Science Technology and
Medicine, Prince Consort Road, London SW7 2BW, UK\\ $^4$School of
Physics and Astronomy, Cardiff University, 5 The Parade, Cardiff CF24
3YB, UK}

\begin{document}

\date{Accepted. Received;}

\pagerange{\pageref{firstpage}--\pageref{lastpage}} \pubyear{2005}

\maketitle

\label{firstpage}

\begin{abstract}

Bright galaxy--galaxy strong lenses are much more powerful than lensed
quasars for measuring the mass profiles of galaxies, but until this
year only a handful have been known. Here we present five new
examples, identified via the optimal line--of--sight gravitational
lens search strategy applied to luminous red galaxies in the Sloan
Digital Sky Survey (SDSS).  Our survey largely complements a similar
survey by Bolton et al., who recently presented several new
lenses. The lensed background galaxies are selected from the SDSS
spectra via the presence of narrow emission line signatures, including
the [OII]~$\lambda\lambda$3726,3729, H$\beta$ and
[OIII]~$\lambda\lambda$4960,5008 lines, superposed on the spectra of
the bright, intervening, deflector galaxies. Our five confirmed new
systems include deflector galaxies with redshifts $z=0.17-0.28$ and
lensed galaxies with redshifts $z=0.47-1.18$. Simulations of
moderately deep (few orbits) HST--ACS imaging of systems such as
these, where the lensed source is brighter than $r\sim 23$, are
presented. These demonstrate the feasibility of measuring accurately
the inner slope of the dark matter halo to within an uncertainty
$\sigma(\gamma)\sim0.1$, the dark matter fraction within the Einstein
radius, and the mass--to--light ratio of the stars alone,
independently of dynamical measurements.  The high success rate of our
search so far, $>60\%$, and the relatively modest observational
resources necessary to confirm the gravitational lens nature of the
candidates, demonstrate that compilation of a sample of $\sim$100
galaxy-galaxy lenses from the SDSS is readily achievable, opening up a
rich new field in dark matter studies.

\end{abstract}


\section{The OLS--lens survey}

This is the second paper in a series presenting the discovery of new
strong galaxy--galaxy lenses, i.e.  systems where a background galaxy
is multiply--imaged by a foreground galaxy.  The motivation for this
work is the opportunity such lenses give for accurately measuring the
projected mass profiles in the lenses, of both the baryons and the
dark matter, with greater precision than has been achieved by other
methods.  Analysis of galaxy--galaxy lenses requires models for the
mass distribution in the lens, and the flux distribution in the
unlensed source (e.g.  Wallington, Kochanek \& Narayan 1996; Warren \&
Dye 2003; Suyu et al. 2006).  The resulting model image is matched to
the observed image, which provides typically hundreds of constraints
because the highly stretched image extends over many resolution
elements.  By contrast, for lensed quasars (i.e.  point sources,
producing point images) the fluxes cannot be used straightforwardly,
because of the possibility individual images are amplified by
microlensing, leaving only the positions of just two or four images as
constraints.  This explains why galaxy--galaxy lenses are much more
powerful than lensed quasars.

By comparison with gravitational lensing, dynamical methods typically
rely on a number of assumptions, such as that the system is relaxed
and is spherically symmetric.  The utility of galaxy--galaxy lenses
for measuring galaxy mass profiles and the advantages of lensing over
dynamics are illustrated by two different analyses of the HST WFPC2
image of the Einstein ring B0047--2808.  The first, by Koopmans \&
Treu (2003), is principally a dynamical analysis, using the measured
radial profile of the stellar velocity dispersion, supplemented by the
constraint provided by the size of the Einstein ring, which measures
the total enclosed projected mass.  The second analysis, by Dye \&
Warren (2005), is a pure lensing analysis, but uses all the
(flux$+$positional) information in the image.  In both analyses the
mass in the lens was assumed to comprise two components; a baryonic
component with mass profile the same as the light profile, of unknown
mass--to--light ratio $\Psi$, embedded in an elliptical non--baryonic
dark--matter halo, with unknown central power--law exponent $\gamma$.
The lensing analysis produced much smaller uncertainties in the
parameters than the dynamical analysis (Fig.  3 in Dye \& Warren,
2005).\footnote{Some caveats to the pure lensing analysis are noted at
  the end of this section.} The baryonic fraction of the projected
mass within the Einstein radius measured in the lensing analysis is
$65^{+10}_{-18}\%$ at $95\%$ confidence.  The size of the uncertainty,
from a single system, is similar to that achieved by Rusin, Kochanek
\& Keeton (2003) in their statistical analysis of 22 lensed quasars.
This further emphasises the advantage of galaxy--galaxy lenses over
lensed quasars. In fact the WFPC2 image of B0047--2808 is of
relatively low signal-to-noise ratio.  As we show below, future
observations of this and other systems with the HST--ACS High
Resolution Camera (HRC), which has higher throughput and better
sampling, will be substantially more powerful.

The number of galaxy--galaxy lenses known where the lensed source is
reasonably bright, e.g.  $r<23$, and therefore suitable for ACS
imaging, was until now very small. The situation has been transformed
in the past few months, and the sample of suitable systems has
increased by an order of magnitude. In the first paper in this series
(Willis, Hewett \& Warren 2005, hereafter WHW05) we listed the three
previously known examples (Warren et al.  1996; Bolton et al.  2005;
Cabanac et al.  2005) and announced the discovery of two new bright
galaxy--galaxy lenses. Smith et al. (2005) have provided a further
example.  Our survey is based on the `optimal line--of--sight' (OLS)
strategy described by Hewett et al.  2000, i.e. targeting those
galaxies that are most likely to lens background sources.  In the
current paper we present five more new bright galaxy--galaxy lenses.
All seven new lenses from our survey were identified as candidates by
searching for anomalous emission lines \---\ typically [OII]3727,
H$\beta$4861 or [OIII]5007 emission \---\ in the spectra of luminous,
bulge--dominated galaxies, in the spectroscopic database of the Sloan
Digital Sky Survey (SDSS).  The SDSS Luminous Red Galaxy (LRG;
Eisenstein et al. 2001) spectroscopic sample forms the bulk of the
spectroscopic targets\footnote{In practice we analyse the spectra of
all SDSS galaxy sources with redshifts satisfying $z>0.25$ for
anomalous emission features. The $z>0.25$ sample is dominated
numerically by LRGs.}.  Full details of our survey (hereafter the
OLS--lens survey) will be provided in a future paper (Hewett et al.
in preparation).  A similar survey is being undertaken by Bolton et
al.  (2004), who list 49 candidates. The two surveys are largely
complementary, and are compared and contrasted in WHW05. Immediately
before submission of this paper, Bolton et al. (2006) announced the
remarkable discovery of 19 new galaxy--galaxy lenses, confirmed by their
morphology in HST--ACS snapshot images. We have checked that none of
our new lenses are in their list.

Our methodology for confirming spectroscopically--selected candidates
is firstly to obtain deeper broadband images than the original SDSS
exposures (54s on a 2.5m telescope), typically 10min exposures in the
$r$ band on a 4m telescope.  Fitting of smooth model galaxy profiles
is used to subtract the image of the deflector galaxy in order to
search for images of the source responsible for the emission line.  In
the majority of cases we find either one or more images within
$2\arcsec$ of the galaxy centre.  But at this stage, even in good
seeing conditions, $<0.8\arcsec$, the results are usually ambiguous as
to whether the system is a gravitational lens, or a chance
superposition, or a peculiar galaxy core.  To confirm that the system
is a gravitational lens we require the spectroscopic detection of two
images of the emission line on either side of the spectrum of the
deflector galaxy.  The $r$ band imaging provides the information to
maximise the chances of success, by aligning the slit along the line
from the galaxy centre to the putative brightest lensed image.  So
far, following these procedures, we have confirmed seven lenses
through observations of just 11 candidates.  The solitary candidate
that shows no evidence for lensing in an $r$ band image of similar
quality to those in Figure 2, is SDSS J212628.78+111808.0,
$z_{gal}=0.34$ and it is likely that the putative 
[OII]~$\lambda\lambda$3726,3729
emission line detection at $7061\,$\AA \ is spurious.  Three
additional candidates show evidence for the presence of extended
morphologies of the background emission line source but suitable
spectroscopic observations remain to be undertaken.  Our success rate
(and that reported by Bolton et al., 2006) demonstrates that the SDSS
DR4 spectroscopic sample (Adelman--McCarthy et al., 2005) contains
many tens of galaxy--galaxy lenses that are easy to identify.  This
resource can provide an increase in the number of confirmed
galaxy--galaxy lenses by over an order of magnitude, such that the
number of known galaxy--galaxy lenses could rival the number of known
lensed quasars.  Given the potential of galaxy--galaxy lenses to
determine the mass distribution within massive galaxies, such surveys
are an important development in the field of gravitational lensing.

Our final goal is to measure the mass profiles of a substantial sample
of deflector galaxies through the lens inversion of deep ACS images of
confirmed systems brighter than $r\sim23$. We have undertaken
end-to-end modeling of ACS observations of galaxy--galaxy lenses, in
order to determine the feasibility of measuring accurate mass profiles
for such a sample in a reasonable number of orbits, and to decide
between the ACS--HRC and ACS Wide Field Camera (WFC, higher
throughput, larger pixels; Gonzaga et al. 2005). We employ the
semi--linear inversion technique of Warren and Dye (2003), and Dye and
Warren (2005). An example of the modeling is shown in
Fig. 1. Successive rows illustrate the input model, and the results
for integrations of 1 orbit with the WFC, 1 orbit with the HRC, and 5
orbits with the HRC. The middle column shows the source (either input,
or reconstructed by the inversion algorithm), and the left--column
shows the image, with pixel size and noise characteristics appropriate
for the selected instrument and integration time. For this simulation
the integrated brightness of the lensed source was $r=22$, the axis
ratios ($q=b/a$) of the baryonic and dark matter components were
$q_b=0.69$, and $q_d=0.8$, and external shear was not modeled. The
right--hand column illustrates the resulting confidence contours on
the two most interesting parameters of the mass model, namely
$\gamma$, the inner slope of the dark matter mass distribution, and
$\Psi$, the restframe B--band mass--to--light ratio of the
baryons. The input values of these two parameters were 0.85 and 3.0
respectively. The points to note are, firstly, that despite the lower
throughput the uncertainties are smaller for the HRC than the WFC for
the same integration time, as a consequence of the better sampling,
and, secondly, that the uncertainties shrink approximately with the
square root of the number of orbits. More quantitatively, the $95\%$
confidence range on the inner slope are $\gamma<1.44$ for 1 orbit with
the WFC, $\gamma<1.29$ for 1 orbit with the HRC, and
$0.51<\gamma<1.00$ for 5 orbits with the HRC. This simulation
demonstrates that for a relatively bright galaxy--galaxy lens it is
feasible to measure the inner slope of the dark matter mass
distribution to an accuracy $\sigma(\gamma)\sim0.1$ with just a few
orbits integration. Similarly tight constraints on $\Psi$, and on the
dark matter mass fraction inside the Einstein radius are obtained.

\begin{figure*}
\caption{Application of the semi-linear method to simulated Einstein
ring images.  The top row shows the input source surface brightness
distribution (centre) and the noiseless, unsmeared lensed image
(left). The lens model comprises a baryonic Sersic profile with mass
to light $\Psi=3$ and a generalised NFW dark halo with slope
$\gamma=0.85$. Successive rows show, from left to right; 1) The
simulated ring image, particular to the ACS camera and number of
orbits as labeled, 2) The reconstructed source, 3) The 68\%, 95\%,
99\% and 99.9\% marginalised confidence limits on the best-fit lens
parameters $\Psi$ and $\gamma$. The dot in each of the right hand
panels indicates the input parameter values of the simulated deflector.}
\label{acs}
\end{figure*}

The above analysis rests on a number of simplifying assumptions,
including that the central dark-matter density profile is of power-law
form, that the dark-matter isodensity contours have constant
ellipticity, that external shear is negligible, and that a constant
M/L for the baryons is appropriate. Relaxing any of these assumptions
could introduce strong correlations between parameters, and so
increase the uncertainties. Fortunately the new HST-ACS data (e.g. Bolton et
al., 2006) are of much higher S/N than the system analysed by Dye and
Warren (2005), and will allow more elaborate mass models to be
explored. It is worth emphasising that the analysis of galaxy-galaxy
lenses is in its infancy, and deep images of a large number of systems
will allow exploration of the relative value, in providing constraints
on the mass profile, of different image configurations (due to the
variables of source impact parameter, lens ellipticity, and source
size, which determines the ring thickness).

In the next section we describe the follow-up imaging (\S2.1) and
spectroscopic (\S2.2) observations of the five new lenses, and then
discuss the evidence that each of the systems is a gravitational
lens. In \S3 we provide a brief summary.

\section{Follow--up imaging and long--slit spectroscopy of the
  candidate sample} 

In the following discussion all quoted $r$ and $i$ magnitudes refer to
the SDSS filter set, and are on the AB system.

\subsection{Imaging}

Deeper red--band observations of the five lens systems presented in
this paper were performed during two separate observing runs.  Table 1
provides details of the imaging observations of the five new systems.
Successive columns list the object name, the J2000 coordinates, the
telescope+camera combination, the seeing, the filter, and the
integration time.  The systems J1246+0440 and J1446--0248 were imaged
in non--photometric conditions using the Prime Focus Imaging Camera
(PFIP {\it (sic)}), with the Harris red filter, hereafter $R$, at the
4m William Herschel Telescope (WHT) on 2005 May 16 and 13
respectively.  The systems J1150+1016, J2156+1204 and J2231--0849 were
observed with the European Southern Observatory (ESO) Multi-Mode
Instrument (EMMI), with the Gunn red filter, hereafter $r_G$, at the
3.5m ESO New Technology Telescope (NTT) during 2005 June 6--7, in
clear conditions.  The data were processed employing standard CCD
reduction techniques.  For both filters $R$ and $r_G$ we computed the
photometric zero point, and colour term to convert to $r$ using the
SDSS measured $r$ and $i$ magnitudes of stars in each frame.  Then,
using the SDSS measured $r-i$ colour of each target galaxy, we
calibrated the counts in the galaxy to the $r$ system.  Note that
photometric conditions are not required for calibration performed in
this way.

\begin{table*}
\caption{Details of the imaging observations of the five systems}
\label{tab_radec}
\centering{
\begin{tabular}{lcccccc}\hline
Lens & RA & Dec. & Telescope$+$Camera & Seeing & Filter & Exposure time (s) \\ \hline
J1150+1016 & 11 50 17.016 & $+$10 16 52.94 & NTT/EMMI & $1.3\arcsec$ & $r_G$ & $3 \times 300$ \\
J1246+0440 & 12 46 45.610 & $+$04 40 25.00 & WHT/PFIP & $1.1\arcsec$ & $R$ & $3 \times 300$ \\
J1446--0248 & 14 46 02.630 & $-$02 48 27.88 & WHT/PFIP & $0.8\arcsec$ & $R$ & $2 \times 300$ \\
J2156+1204 & 21 56 27.149 & $+$12 04 11.69 & NTT/EMMI & $0.8\arcsec$ & $r_G$ & $2 \times 300$ \\
J2231--0849 & 22 31 08.018 & $-$08 49 25.50 & NTT/EMMI & $0.7\arcsec$ & $r_G$ & $3 \times 300$ \\  \hline
\end{tabular}
}
\end{table*}

\begin{figure*}
\caption{Follow--up $r_G$/$R$--band images for each lens system (see
Table \ref{tab_radec} for details). For each system the left and right
panels show the image before and after subtraction of the deflector
galaxy surface brightness profile (see text). The centroid of the
galaxy surface brightness distribution is indicated in each
unsubtracted frame by a white cross. In each case the centroid of the
subtracted galaxy image is located between the residual images. The
orientation of the image of each system has been selected to match the
corresponding two--dimensional spectral image of each system displayed
in Fig. \ref{5spec}. The exact slit/image rotation angles for each
system are shown in Table \ref{tab_lens_values}.}
\label{5image}
\end{figure*}

To look for evidence of lensing (arcs or multiple images), the surface
brightness distribution of the deflector galaxy was modeled and
subtracted.  We used the elliptical Sersic model for the
surface--brightness profile, convolved with the point spread function,
defined by stars in the frame. The model has the following seven free
parameters: x, y, orientation, ellipticity, half-light radius
$r_{0.5}$, surface brightness $\Sigma_{0.5}$ at $r_{0.5}$, and Sersic
index $n$. The fitting was iterated, successively improving the
masking of any underlying images revealed. Fuller details of the
fitting procedure are provided in Wayth et al. (2005). While the
subtraction in each case is visually satisfactory and therefore
suitable for the purposes of searching for lensed images, in most
cases the $\chi^2$ of the fit is formally unsatisfactory. Table
\ref{sersicfits} provides the best--fit values of the three parameters
$r_{0.5}$, $\Sigma_{0.5}$, and $n$ for each system, together with the
computed total magnitude. The remaining entries in Table 2 are the
total magnitude of the source image (see below), the source redshift
(from \S2.2), and the redshift and velocity dispersion (if provided)
of the lens galaxy, taken from the SDSS database.

We find very good agreement between our lens--galaxy total magnitudes
and the SDSS de Vaucouleurs--model total magnitudes, in the following
sense. For values of $n>4$ our total magnitudes are brighter than the
SDSS model magnitudes, as would be expected, because at large radii,
beyond the fitted region, our surface brightness profiles fall off
less steeply than the de Vaucouleurs model. Where our measured
profiles have $n<4$ the opposite is true. Plotting the magnitude
difference between our model and the SDSS model against $n$, and
fitting a linear relation, the offset at the value $n=4$ is only
0.04mag.

The images of the five fields both before and after subtraction are
displayed in Fig. \ref{5image}. Each image is oriented to match the
position angle of the spectroscopic observation such that the slit
lies horizontally in these images. In each case, subtraction of the
galaxy surface brightness profile revealed the presence of additional
sources within 1--2\arcsec\ of the galaxy centroid, i.e.  within the
typical Einstein radius expected if the galaxy is acting as a massive
deflector\footnote{For example, for a singular isothermal sphere at
$z=0.4$, of one--dimensional velocity dispersion
$\sigma_v=220$~km\,s$^{-1}$, the angular diameter of the Einstein ring
is 1\farcs5 and 2\farcs2 for source redshifts of $z=1$ and $z=4$
respectively.}.  We note that the residual image features displayed in
Fig. \ref{5image} are, in some cases, not unlike features anticipated from the
subtraction of an imperfect parameterized model (i.e., to the extent
that the lens galaxies are not exactly Sersic ellipsoids). However,
while we acknowledge the possibility, this explanation is ruled out by
the results of the long slit spectroscopy presented in \S2.2. It is
difficult to estimate the brightness of the sources, because they lie
at small radii, where the residuals from the galaxy subtraction are
greatest. Nevertheless, under the lens hypothesis, most of the flux
will lie in an annulus close to the Einstein radius, whereas the worst
galaxy residuals will occur at the smallest radii. With this in mind
we adopted a simple recipe, integrating the flux in the subtracted
frames in an annulus of radial extent
$0.75\arcsec<r<2.00\arcsec$. These values are entered in Table 2. We
have not attempted to estimate the uncertainty on these values, which
are dominated by systematic errors associated with the galaxy
subtraction. The average estimated source total magnitude is $r=22.1$,
which is encouraging in the light of the simulations presented in
Fig. 1.

We now discuss each image in turn.,

{\bf J1150+1016:} The subtracted image shows a promising
incomplete--ring structure like a horseshoe.  At the same time there
are relatively large +/- residuals near the galaxy centre, and we
found that the flux distribution around the candidate ring was
somewhat sensitive to the exact masking of the source employed. This
is as expected given the rather poor seeing (Table 1) and the relative
faintness of the source (Table 2). Therefore while we might expect a
higher--resolution image of the system to present a somewhat different
morphology, the image is nevertheless very suggestive of lensing, and
we oriented the slit E--W to maximise the chance of detecting the
emission line on either side of the lensing galaxy.

{\bf J1246+0440:} The subtraction reveals a pair of faint images
either side of the galaxy centre, consistent with the lensing
hypothesis. The flux ratio between the two residual images was rather
sensitive to the exact mask used. The positive residuals are
substantially larger than the negative residuals. Therefore we were
confident that both images are real, and aligned the slit to include
both.

{\bf J1446-0248:} We see two relatively bright residual images, either
side of the target galaxy, consistent with the lensing hypothesis. We
aligned the slit to include both images.

{\bf J2156+1204:}  This image is rather complicated.  There is an elongated
image, resembling an inclined spiral at $3\arcsec$ to the S, and a smaller image
$1\arcsec$ further to the S.  In our spectrum we detect continuum from the
latter, and an emission line near $7233$\AA\, from the former.  The line might
be [OII]~$\lambda\lambda$3726,3729 or H$\alpha$ $\lambda$6565, but in either
case the object is not at the redshift of the target galaxy or of the source.
The measured source magnitude quoted in Table 2 will be brighter than the
reality because some flux from this object will be included in the measurement
annulus.  Closer in, a faint residual structure reminiscent of the horshoe seen
for J1150+1016 is weakly visible.  Similar remarks about the morphology of the
structure apply here i.e.  it is suggestive of lensing, but a higher-resolution
image may look somewhat different.  We aligned the slit NS, as the image
indicates flux from the source on either side of the target galaxy along this
axis.

{\bf J2231-0849:} The residual image for this source shows a
remarkably bright ring. Varying the region masked made little
difference to this conclusion, and we considered it highly likely that
this is a bright example of an Einstein ring. We aligned the slit to
intersect the brightest part of the ring.

\begin{table*}
\caption{Measured properties of the deflector and source galaxies in
each system}
\label{sersicfits}
\centering{
\begin{tabular}{lccccc}\hline
                    & J1150+1016 & J1246+0440 & J1446--0248 & J2156+1204 & J2231--0849 \\ \hline
$r_{0.5}$ (arcsec)  & $2.02\pm0.05$ & $1.70\pm0.04$ & $3.67\pm0.11$ &
  $4.88\pm0.51$ & $1.57\pm0.01$ \\  
$\Sigma_{0.5}$ ($r$ mag/arcsec$^2$) & $22.68\pm0.05$ & $21.97\pm0.05$
  & $23.90\pm0.06$ & $24.91\pm0.22$ & $21.67\pm0.02$ \\   
$n$                 & $5.08\pm0.11$ & $5.36\pm0.12$ & $5.66\pm0.11$ &
  $7.71\pm0.36$ & $3.00\pm0.03$ \\   
lens $r$       & 17.64 & 17.27 & 17.51 & 17.73 & 17.45 \\  
source $r$       & 22.8 & 23.9 & 21.9 & 22.3 & 19.7 \\  
$z_s$               & 0.801 & 0.465& 0.697 & 1.176 & 0.597 \\
{[OII]}3727 flux  & $54 \pm 6$ & $52 \pm 6$ & $26 \pm 4$ & $46 \pm 5$ & $89 \pm 8$ \\
($\times 10^{-17}$ ergs s$^{-1}$ cm$^{-2}$) &&&&&\\
$z_l$               & 0.284 & 0.169 & 0.206 & 0.257 & 0.175 \\
$\sigma_v$ (kms$^{-1}$) & ... & $217 \pm 19$ &  ...  &  ...  & ... \\  \hline
\end{tabular}
}
\end{table*}
\begin{table*}
\caption{Details of the spectroscopic observations of each system, and
the measured parameters for the detected [OII]~$\lambda\lambda$3726,3729
lines. The designation image ``A'' is assigned to the brightest of the
two source images measured in emission line flux. The 
measured image separations are affected by systematic uncertainties associated
with the subtraction of the galaxy spectrum. We use a nominal
uncertainty of 0\farcs1.}
\label{tab_lens_values}
\begin{tabular}{lccccc}
\hline
& J1150+1016 & J1246+0440 & J1446--0248 & J2156+1204 & J2231--0849 \\
\hline
Slit position angle ($^{\circ}$EofN) & 90 & $-10$ & 100 & 0 & 135 \\
Exposure time (s)    & $3\times1800$  & $3\times 1800$ & $4\times 1800$ & $3\times 1800$ & $4\times 1200$ \\
Image separation (\arcsec)   & $2.62 $ & $1.83 $ & $1.73 $ & $2.12 $ & $2.16 $ \\
$|\rm Galaxy - A|$ (\arcsec) & $1.00 $ & $1.23 $ & $1.12$   & $1.06 $ & $1.36$ \\
$|\rm Galaxy - B|$ (\arcsec) & $1.63$ & $0.60$ & $0.64$   & $1.06$ & $0.80 $ \\
Flux ratio A/B               & 1.92            & 2.83            & 3.60            & 1.12            & 1.11            \\
SNR(B)                       & 19.5            & 15.8             & 5.0             & 10.3            & 20.3            \\
\hline
\end{tabular}
\end{table*}

\begin{figure*}
\caption{Two--dimensional reduced spectroscopic frames of each lens
 system. For each system, the upper and lower panels correspond to
 different stages of the analysis. In each case, the spectral axis is
 oriented vertically (with lower wavelengths at the bottom of the
 panel) and the spatial axis is oriented horizontally: the displayed
 dimension of each panel is approximately 24\arcsec $\times$
 60\AA. {\em Upper panel:} Before sky subtraction, {\em lower panel:
 } after continuum subtraction of sources, and convolution with a
 Gaussian of $\sigma=1$ pixel.}
\label{5spec}
\end{figure*}

\subsection{Spectroscopy}

We followed the same procedures for the spectroscopy as in WHW05,
i.e. we oriented the slit at a position angle suggested by the
imaging, and then after data reduction and sky subtraction, we
subtracted the spectrum of the deflector galaxy, and searched for
paired images of an emission line on either side of the position of
the galaxy, at the wavelength of the emission line(s) detected in the
original SDSS spectrum. Finally we extracted near--optimal spectra of
the sources to search for additional lines to corroborate the
redshifts. Only brief details of the procedures for subtracting the
galaxy spectrum, and extracting the source spectrum are provided
here. The reader is referred to WHW05 for fuller details.

Spectroscopic observations were performed during 2005 June 6--7 using
the Red Image Low Dispersion (RILD) configuration of the EMMI
instrument at the ESO NTT.  Systems J1150+1016, J1446--0248,
J2156+1204 and J2231--0849 were observed using EMMI Grism \#6 and a
1\farcs0 slit, providing spectral coverage from 5800 to 8500\AA, at a
resolving power $\sim 1500$.  System J1246+0440 was observed with EMMI
Grism \#5 and a 1\farcs0 slit, providing spectral coverage from 4000
to 7000\AA, at a resolving power $\sim 1100$. The spatial scale at
each configuration was 0\farcs33 pix$^{-1}$. Integrations were
typically split into either three or four exposures of 1800s each. The
slit position angle and exposure time used for each system are
provided in Table \ref{tab_lens_values}.  The typical atmospheric
seeing over the two observing nights was 0\farcs8 to 1\farcs0 although
observations of J1246 were performed in 1\farcs2 seeing. Conventional
procedures were followed for bias subtraction and flatfielding.  The
multiple frames for each target were then averaged, employing a
sigma--clipping algorithm in order to remove cosmic ray events.

Small sections of the final frames for each system are shown in Fig.
\ref{5spec}.  For each system, the upper panel shows the reduced two
dimensional frame prior to sky subtraction. After sky subtraction, a
low order cubic spline function was fit up each column. This is
effective in subtracting the spectrum of the lens galaxy (at
wavelengths away from the strongest absorption lines). The
resulting frames, convolved with a Gaussian of $\sigma=1$ pixel, for
display purposes only, are provided in the lower panels for each
system. All analysis was undertaken on the unconvolved frames. All
systems show an emission line at the expected wavelength, split on
either side of the lens galaxy. In each case we conclude that the
emission line seen is [OII]~$\lambda\lambda$3726,3729, and
consequently the detected object lies at a redshift larger than the target
galaxy. These two facts combined are sufficient to conclude that each
system is a gravitational lens.  We note that two dimensional long
slit spectroscopy provides only limited constraints on the detailed
image morphology of the lensed emission line source. The important
point however, is that resolution of the emission line source into
multiple components provides compelling evidence for the lensing
hypothesis. We discuss each system in turn below.

In Table 3 we summarise the characteristics of the image configuration,
as measured in the spectrum, for each system, listing the image
splitting, the impact parameters of the primary (A, i.e. brightest)
and secondary (B) images, the flux ratio, and the signal-to-noise
ratio (SNR) of the detection of the fainter image. In some cases the
SNR of the detections of the two images is less than in the $r$--band
images. Nevertheless, because the contrast of the line over the
spectrum of the galaxy is much greater than in the $r$--band images,
the systematic errors from subtraction of the galaxy will be
smaller. Therefore we consider that the relevant quantities summarised
in Table 3 are more reliably derived from the spectroscopic frames
than from the $r$--band images.

\begin{figure*}
\includegraphics[width=80mm]{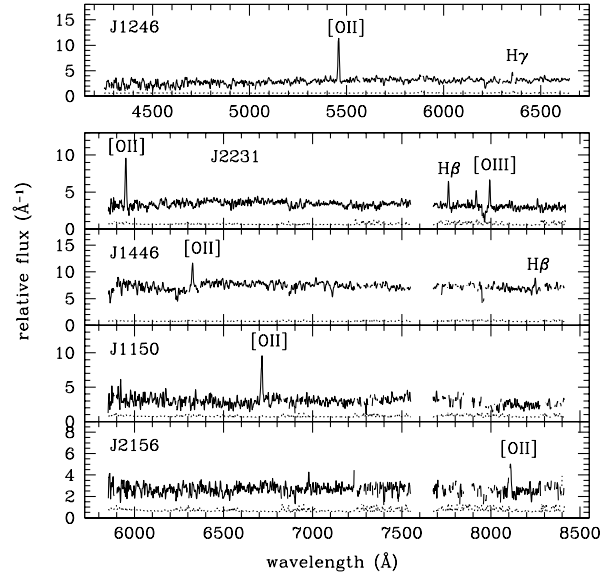}
\caption{Extracted spectra of the source galaxies in each system. The
spectra of confirmed lenses are presented in order of increasing
source redshift from the top to the the bottom panels. Prominent
emission features are indicated. Note that J1246 was observed using a
different spectrograph configuration to the other lens systems and is
thus presented in a separate panel.}
\label{5spec_1d}
\end{figure*}

We extracted one dimensional spectra of the deflector galaxy and the
bright and faint images of the lensed emission--line galaxy in each
system using the procedure described in WHW05. The procedure models
each component (deflector galaxy plus images A and B) on the reduced
two--dimensional spectral frame as a Gaussian spatial profile of
specified Full--Width at Half--Maximum (FWHM).  The relative spatial
separation of the components and the flux ratio between image A and B
are fixed.  The fitting procedure computes a best--fitting system
centroid and vertical normalisation for the deflector galaxy and image
A component by minimising the $\chi^2$ statistic.  The resultant
two--dimensional model provides the relative contribution of the
deflector galaxy, image and A and B at each spatial pixel, permitting
straightforward extraction of the data for each source. The extracted
spectra of the sources are plotted in Fig. \ref{5spec_1d}.

We now discuss each system in turn:

{\bf J1150+1016:} The emission line is clearly resolved into multiple
spatial components.  The evidence that the line is
[OII]~$\lambda\lambda$3726,3729 is the observation that the lines are
double peaked at a separation corresponding to the [OII] doublet.  The
absence of other emission lines in the spectrum corroborates this
identification but the definitive evidence comes from the detection of
very strong [OIII]~$\lambda\lambda$4960,5008 emission in the original
SDSS spectrum.  The measured flux ratio for this system is at first
sight peculiar.  For a simple model of a singular isothermal sphere
(SIS) and a point source, for images A and B at impact parameters $a$
and $b$, one predicts a flux ratio $a/b=0.61\pm0.20$, inconsistent
with the observed value of 1.92.  However this simple rule can be
completely misleading for an extended source lensed by an elliptical
galaxy.  The arc configuration suggested by the $r$--band image can be
produced by an extended source straddling the astroid caustic.

{\bf J1246+0440:} The emission line is clearly resolved into multiple
spatial components.  The emission line is not resolved (recall that
this spectrum has lower resolution than the others).  The evidence
that the line is [OII]~$\lambda\lambda$3726,3729 is the detection of
the H$\gamma$ emission line (just blueward of a sky emission line near
$\lambda$6360),
visible as the blip at the red end of the spectrum.  The NTT spectrum
thus directly confirms the multiple emission line detections,
including H$\beta$ and [OIII]~$\lambda\lambda$4960,5008 in the
original SDSS spectrum.  The $r$ image indicates a simple two--image
configuration.  Therefore here the SIS model may be a reasonable
approximation.  The predicted flux ratio $a/b=2.1\pm0.4$ is in
agreement with the measured value of 2.83.  From the $r$ band image we
obtain $a/b=2.6$. The measured image separation is also consistent
with the prediction of a SIS model and the SDSS deflector velocity
dispersion shown in Table 2 (approximately $1\farcs65 \pm 0\farcs3$).

{\bf J1446--0248:} The emission line is resolved into multiple spatial
components, but the significance of the detection of the weaker image,
S/N$=5$, is the weakest of all the candidates, and the image is only
visible in Fig.  3 as a faint extension to the left.  The evidence
that the line is [OII]~$\lambda\lambda$3726,3729 is the observation
that the lines are double peaked at a separation corresponding to the
[OII] doublet, corroborated by the detection of H$\beta$ at the
correct wavelength.  The original SDSS spectrum also clearly shows the
[OIII]~$\lambda\lambda$4960,5008 doublet.

{\bf J2156+1204:} The emission line is clearly resolved into multiple
spatial components. The evdence that the line is
[OII]~$\lambda\lambda$3726,3729 is the observation that the lines are
double peaked at a separation corresponding to the [OII] doublet, in
agreement with the profile of the line in the original SDSS
spectrum. No other emission lines are expected in either spectrum
given the redshift of $z_l=1.176$.

{\bf J2231--0849:} The emission line is clearly resolved into multiple
spatial components. The evdence that the line is
[OII]~$\lambda\lambda$3726,3729 is the detection of the emission lines
H$\beta$ and [OIII]~$\lambda\lambda$4960,5008 at the correct
wavelengths (Fig. 4) in agreement with the features seen in the
original SDSS spectrum.

\section{Summary}

The optimal line--of--sight selection strategy applied to the SDSS
spectroscopic database provides an efficient means to identify
galaxy--galaxy lenses consisting of massive bulge--dominated
deflectors, $0.1 < z < 0.5$, and background lensed star--forming
galaxies, $0.3 < z < 1.2$.  Five new systems are presented, bringing
our sample of confirmed lenses to seven.  Our sample extends the upper
redshift limit of the lensed galaxies from $z\simeq 0.8$ in the Bolton
et al.  (2006) sample to $z \simeq 1.2$.  Further galaxy--galaxy
lenses from very recent observing runs will be presented in a
subsequent paper and many tens of high-probability candidates remain
to be observed.  In parallel with the work of Bolton et al. (2005,
2006) we have demonstrated the feasibility of detecting as many as 100
SDSS galaxy--galaxy lens, offering the prospect of making dramatic
advances in our understanding of the mass distributions in massive
galaxies. Our own simulations demonstrate that the most effective
observing strategy for providing tight constraints on the inner mass
profiles of individual deflector galaxies is to obtain relatively high
SNR imaging of the systems with the ACS--HRC on Hubble Space
Telescope.

\section*{acknowledgements}

Funding for the Sloan Digital Sky Survey (SDSS) has been provided by
the Alfred P. Sloan Foundation, the Participating Institutions, the
National Aeronautics and Space Administration, the National Science
Foundation, the U.S. Department of Energy, the Japanese
Monbukagakusho, and the Max Planck Society. The SDSS Web site is
http://www.sdss.org/.

The SDSS is managed by the Astrophysical Research Consortium (ARC) for
the Participating Institutions. The Participating Institutions are The
University of Chicago, Fermilab, the Institute for Advanced Study, the
Japan Participation Group, The Johns Hopkins University, Los Alamos
National Laboratory, the Max-Planck-Institute for Astronomy (MPIA),
the Max-Planck-Institute for Astrophysics (MPA), New Mexico State
University, University of Pittsburgh, Princeton University, the United
States Naval Observatory, and the University of Washington.

NM wishes to thank the Overseas Research Students Awards Scheme, the
Cambridge Commonwealth Trust, and the Dr. John Taylor Scholarship from Corpus
Christi College for their generous support.

\bsp

\label{lastpage}

\end{document}